\documentclass[sort&compress
    ,final            
  ]
  {aipproc}

\layoutstyle{8x11single}

\usepackage{amsfonts}
\usepackage{eucal}
\usepackage{bm}
\usepackage{amssymb}
\usepackage{amsmath}

\begin{document}

\title{Nonlocal Gravity}


\author{Bahram Mashhoon}{address={Department of Physics and Astronomy, University of Missouri, Columbia, MO 65211, USA}
}


\def\a{\alpha}
\def\b{\beta}
\def\g{\gamma}
\def\vt{\vartheta}
\def\d{\delta}
\def\vta{\vartheta}
\def\ve{\varepsilon}
\def\wa{\widetilde{\alpha}}
\def\vp{\varphi}

\def\stareq{\stackrel{*}{=}}

\def\S{\Sigma}
\def\@{\partial_}

\def\chris#1#2#3{#1\brace #2 #3}
\def\negenspace{\kern-1.1em}\def\quer{\negenspace\nearrow}

\def\sqr#1#2{{\vcenter{\hrule height.#2pt\hbox{\vrule width.#2pt
height#1pt \kern#1pt \vrule width.#2pt}\hrule height.#2pt}}}
\def\square{\mathchoice\sqr64\sqr64\sqr{4.2}3\sqr{3.0}3}
\def\hatsquare { {\hat\sqcap\!\!\!\!\sqcup} }

\def\bfx{\mathbf{x}}
\def\bfxd{\mathbf{x'}}

\def\bfy{\mathbf{y}}
\def\bfyd{\mathbf{y'}}

\def\bfz{\mathbf{z}}
\def\bfzd{\mathbf{z'}}

\def\bfr{\mathbf{r}}

\def\bfk{\mathbf{k}}
\def\bfkd{\mathbf{k'}}
\def\bfK{\mathbf{K}}

\def\rhod{\rho_{\rm D}}

\begin{abstract}
The analysis of measurements of accelerated observers in Minkowski spacetime has led to the development of nonlocal special relativity theory. Inertia and gravitation are intimately connected in accordance with the principle of equivalence. We therefore seek a nonlocal generalization of the theory of gravitation such that in the new theory the field equations are integro-differential equations for the local gravitational field. We show that it is possible to develop a nonlocal generalization of Einstein's theory of gravitation via the introduction of a scalar ``constitutive'' kernel in the teleparallel equivalent of general relativity. The resulting nonlocal theory is essentially equivalent to Einstein's theory plus ``dark matter''. That is, nonlocality simulates dark matter by introducing a new source term into general relativity. In the linear approximation for the nonlocal modification of Newtonian gravity, we recover the theoretical basis for the phenomenological Tohline-Kuhn modified gravity approach to the explanation of the astrophysical evidence for dark matter.

\end{abstract}

\maketitle


\section{Introduction}

	A crucial observational aspect of gravitation is the universality of free fall within the classical domain of physics.  In the context of Newtonian theory of gravitation, this experimental result implies the principle of equivalence of inertial and gravitational masses. The deep connection between inertia and gravity is the origin of Einstein's principle of equivalence, which establishes a definite local relationship between acceleration and gravitation~\cite{1}.  Therefore, following Einstein, to develop a theory of the gravitational field, our starting point must naturally be the physics of accelerated observers in Minkowski spacetime. 
	 
	Imagine a global background inertial reference system with Cartesian coordinates $x^{\mu}= (ct, \bfx)$.  Henceforth we use units such that $c=1$, unless specified otherwise.  Consider an accelerated observer following a world line $x^{\mu}(\tau)$ and an orthonormal tetrad frame  $\lambda^{\mu}{}_{(\alpha)}$ carried by the observer along the path such that  $\lambda^{\mu}{}_{(0)}=dx^{\mu}/d\tau$ is the unit timelike vector tangent to the path as well as the four-velocity vector of the observer.  Here $\tau$ is the proper time along the world line. The motion of the local frame of the observer is given by 
\begin{equation}\label{1}
\frac{d\lambda^{\mu}{}_{(\alpha)}}{d\tau}=\phi_{(\alpha)}{}^{(\beta)}(\tau)~\lambda^{\mu}{}_{(\beta)}\,,
\end{equation}
where $\phi_{(\alpha)(\beta)} = - \phi_{(\beta)(\alpha)}$ is the antisymmetric acceleration tensor of the observer.  In analogy with electrodynamics, $\phi_{(\alpha)(\beta)}$ has components  $(\mathbf{a}, \mathbf{\omega})$, so that $\phi_{(\alpha)(\beta)}$                                                                          can be expressed in terms of the translational acceleration of the observer $(\mathbf{a})$ and the rate of rotation of the spatial frame with respect to a nonrotating (i.e., Fermi-Walker transported) frame $(\mathbf{\omega})$. Out of these spacetime-invariant quantities, one can construct acceleration scales that characterize the variation of the state of the observer.  Once the kinematic aspects of the motion of an arbitrary observer in Minkowski spacetime are established, we turn to the problem of physics in accelerated systems, namely, what accelerated observers actually measure.

	In the standard theory of special relativity, Lorentz invariance is extended to accelerated observers in a pointwise manner via the hypothesis of locality~\cite{2,3,4,5}. That is, an accelerated observer is assumed to be pointwise inertial.  This assumption originates from Newtonian mechanics, where the state of a point particle is characterized only by its position and velocity. For instance, the locality principle implies that the proper time $\tau$ is in fact the time as determined by the accelerated observer (``clock hypothesis'').  This approach is justified for phenomena involving classical point particles and rays of radiation, since in these regimes instantaneous measurements are in principle possible. It generally breaks down, however, for basic field measurements, as these cannot be performed instantaneously.  Indeed, according to Bohr and Rosenfeld, the electric field  $\mathbf{E}(t,\mathbf{x})$      and the magnetic field  $\mathbf{B}(t,\mathbf{x})$        occur in Maxwell's equations as idealizations; only the spacetime averages of these fields have immediate physical significance~\cite{6,7}. While this observation appears innocuous for inertial observers, it acquires the status of a fundamental principle for accelerated observers due to the existence of invariant acceleration scales.  The Bohr-Rosenfeld principle implies that in general field measurements cannot be instantaneous and hence must  depend on the past world line of the observer~\cite{8}. 

	Let $\psi$  be a basic field in the global background inertial frame and let $\hat{\psi} = \Lambda \psi$        be the field measured instantaneously by the infinite set of hypothetical momentarily comoving inertial observers along the world line of an accelerated observer. Here $\Lambda$    is a matrix representation of the Lorentz group.  Suppose that $\hat{\Psi}(\tau)$       is the field actually measured by the accelerated observer. We are interested in the relationship between $\hat{\Psi}$   and $\hat{\psi}$. The most general relation between $\hat{\Psi}$   and $\hat{\psi}$ that is consistent with causality and linearity can be expressed as 
\begin{equation}\label{2}
 \hat{\Psi}(\tau)=  \hat{\psi}(\tau) +  u(\tau - \tau_{0})\int_{ \tau_{0}}^{\tau} \hat{K}(\tau,\tau') \hat{\psi}(\tau')d\tau' \,.
\end{equation}
Here $u(t)$     is the unit step function such that  $u(t) = 0$  for  $t<0$ and $u(t) = 1$ for $t>0$, $\tau_{0}$ is the instant at which acceleration is turned on and the kernel $\hat{K}$     vanishes in the absence of acceleration.  Equation (2) is a Volterra integral relation~\cite{9,10,11} and represents the idea that what is measured at proper time $\tau$ by the accelerated observer is the field $ \hat{\psi}(\tau)$     measured by the instantaneously comoving inertial observer at $\tau$ together with a certain average over the observer's past world line that constitutes the linear memory of past acceleration. 

          For sufficiently low accelerations, the observer can be considered almost inertial and hence the nonlocal part of equation (2) constitutes only a small perturbation in this case. In general, if the intrinsic scale of the phenomenon under observation is very much smaller than the scale of variation of the state of the observer, then the deviation from locality is expected to be negligible. This is indeed the case for most Earth-based observations as $c^{2}/|\mathbf{a}_{\oplus}|\approx 1$ light year and $c/|\mathbf{\omega}_{\oplus}|\approx 28$ astronomical units.  

	To determine the nonlocal contribution to equation (2), we need a physical principle that could form the basis for the calculation of the kernel.  The general phenomenon of spin-rotation coupling furnishes the main idea: No observer can stay completely at rest with a basic radiation field.  This is a generalization for all observers of a well-known result of Lorentz invariance for inertial observers. It follows from a detailed analysis that in general the kernel is directly proportional to acceleration and is given by~\cite{12,13,14,15,16}
\begin{equation}\label{3}
\hat{K}(\tau,\tau')=\hat{k}(\tau')=-\frac{d\Lambda(\tau')}{d\tau'}\Lambda^{-1}(\tau')\,.
\end{equation}                                                                                                                                    
For instance, this is the appropriate kernel for the Dirac field as well as for the electromagnetic gauge field $A_{\mu}(x)$.  The special case of the electromagnetic field $F_{\mu\nu}(x)$       requires a separate treatment~\cite{17}.  

	A nonlocal theory of accelerated observers, first proposed in 1993~\cite{8}, has been further developed in its essential aspects and the main outline of a complete nonlocal special relativity theory has been presented in 2008~\cite{18}.  Some of the observational consequences of the new theory have been worked out, especially regarding spin-rotation coupling~\cite{19}.  In this connection, Bohr's correspondence principle has been employed to show that the nonlocal theory is in better agreement with quantum mechanics than the standard local special relativity theory~\cite{15}. Moreover, the nonlocal theory forbids the existence of basic scalar or pseudoscalar fields.  The nonlocal theory is consistent with all available observational data and its novel consequences should be subjected to direct experimental tests~\cite{19}. 
 
	It is important to emphasize that nonlocal special relativity actually involves local fields $\Psi(x)$       in Minkowski spacetime that satisfy integro-differential field equations.  It turns out that it is not possible to develop a consistent field theory of bilocal fields $\Psi(x,x')$.                          Therefore, the fields are local, but the acceleration-induced nonlocality that originates from the memory of past observer acceleration appears in the circumstance that the local fields satisfy nonlocal field equations~\cite{20}. 
   
	Once a proper nonlocal theory of accelerated systems is on hand, it is natural to seek a nonlocal theory of gravitation due to the basic connection between inertia and gravity.  At first sight, it might seem that one could employ Einstein's principle of equivalence and arrive at nonlocal general relativity in a straightforward manner. It turns out, however, that Einstein's beautiful approach does not work in this case due to the inherently \emph{local} nature of Einstein's principle of equivalence.  For instance, suppose that we start with Einstein's general linear approximation to general relativity theory as a classical spin-2 field in Minkowski spacetime. Nonlocal theory of accelerated systems can then be applied in this case to generate a linear nonlocal spin-2 field theory as determined by accelerated observers in Minkowski spacetime~\cite{21,22}. It is however not clear how to implement in this case Einstein's heuristic pointwise identification of an observer in a gravitational field with a certain accelerated observer in Minkowski spacetime.  Thus a \emph{direct} attempt at a nonlocal generalization of Einstein's theory of gravitation appears to be futile. 
   
	It is possible to arrive at general relativity (GR) from the standpoint of the gauge theories of gravitation.  Indeed, the gauge approach to gravity naturally leads to spacetime theories with curvature and torsion.  There is a spectrum of such theories such that at one end of the spectrum, one has GR based on a Riemannian spacetime manifold with only curvature and no torsion, while at the other end of the spectrum are spacetime theories with torsion and no curvature~\cite{23,24,25,26,27,28}. Of the latter, there is a unique one that is essentially equivalent to Einstein's general relativity: This is the teleparallel equivalent to general relativity (GR$_{||}$). Teleparallelism has a long history; its application to gravitational physics has been considered by many authors starting with Einstein in 1928~\cite{28}.  GR$_{||}$  is a translational gauge theory of gravity and has been recently studied by a number of authors~\cite{29,30,31,32,33,34,35,36,37,38,39,40,41,42,43,44,45,46}.  Spacetime translations form an Abelian group; this fact led Friedrich Hehl to suggest that one should attempt a nonlocal GR$_{||}$ in analogy with the nonlocal electrodynamics of accelerated observers~\cite{47} in order to arrive indirectly at a nonlocal generalization of general relativity.  This fruitful suggestion has been developed in recent papers~\cite{48,49,50}.  The resulting nonlocal theory implies that gravity is nonlocal even in the Newtonian regime. The nonlocally modified Newtonian gravitation appears to provide a natural explanation for the dark matter problem; that is, nonlocality simulates dark matter.
     
	The plan of this paper is as follows.  GR$_{||}$ is briefly described in the following section.  Its nonlocal generalization is then presented and the corresponding field equations are worked out, for the sake of simplicity, in the linear approximation. The nonlocal modification of Newtonian gravity is discussed and its connection with the dark matter problem is emphasized.  We conclude with a brief discussion of these results.

\section{GR$_{||}$}
The arena for the teleparallel equivalent of general relativity is the Weitzenb\"ock spacetime. This gravitational theory can be characterized as a tetrad theory. The gravitational potentials are given by the four coframe 1-forms $\vt^\a=e_i{}^\a dx^i$, where $e_i{}^\a(x)$ is the tetrad field. Here Latin indices range over ${0, 1, 2, 3}$ and refer to holonomic spacetime indices, while Greek indices range over ${\hat{0},\hat{1},\hat{2},\hat{3}}$ and refer to anholonomic tetrad indices. The latter are raised and lowered via the Minkowski metric tensor $\eta_{\a\b}$ = diag$(1,-1,-1,-1)$. The frame vectors $e_\a=e^j{}_\a\partial_j$ are dual to the coframe 1-forms, so that
\begin{equation}\label{4}
e_i{}^\a e^i{}_\b=\d^\a_\b, \quad   e_i{}^\a e^j{}_\a=\d_i^j.  
\end{equation} 

The gravitational field strength is given by $ C^\a:=d\vt^\a$, hence
\begin{equation}\label{5}
C^\a=\frac 12 C_{ij}{}^\a dx^i\wedge dx^j,
\end{equation}
where $C_{ij}{}^\a$ can be expressed as
\begin{equation}\label{6}
 C_{ij}{}^\a=2\partial_{[i}e_{j]}{}^\a\,.
\end{equation}
The spacetime interval is given by $ds^2=g_{ij}dx^i\otimes dx^j$, where
\begin{equation}\label{7}
g_{ij}= e_i{}^\a e_j{}^\b \eta_{\a\b}\,.
\end{equation}
We define a geodesic in this spacetime connecting two fixed events $P'$ and $P$ to be such that the corresponding interval is an extremum; that is,   $\delta\int_{P'}^P ds=0$. This implies that
\begin{equation}\label{8}
\frac{d^2x^i}{ds^2}+{\chris{i}{j}{k}}\frac{dx^j}{ds}\frac{dx^k}{ds}=0\,,
\end{equation}
where the Christoffel symbols are given by
\begin{equation}\label{9}
  {\chris{i}{j}{k}}=\frac 12\,g^{il}\left(g_{lj,k}+g_{lk,j}-g_{jk,l}\right).
\end{equation}

In the Weitzenb\"ock geometry, the frame field is globally teleparallel. This means that the connection 1-form, $\Gamma^{\a\b}=-\Gamma^{\b\a}=\Gamma_i{}^{\a\b}dx^i$, is such that
\begin{equation}\label{10}
  R_\a{}^\b
  :=d\Gamma_\a{}^\b-\Gamma_\a{}^\g\wedge \Gamma_\g{}^\b 
=\frac 12 R_{ij\a}{}^\b dx^i\wedge dx^j=0\,.
\end{equation}
Therefore, one can choose a suitable Cartesian tetrad frame such that 
\begin{equation}\label{11}
  \Gamma^{\a\b}\stareq 0\,.
\end{equation}
The situation here is similar to that of Minkowski spacetime in which the Riemannian curvature vanishes; then, it is possible to choose inertial Cartesian coordinates such that the connection coefficients all vanish globally. Henceforth, a star over the equality sign indicates that the corresponding relationship is only valid in the particular gauge in which equation (11) holds. 

While the frame field is globally parallel and curvature-free, the spacetime torsion is nonzero. In fact, $T^\a:=D\vt^\a$ is simply the gravitational field strength $C^\a$ in the special gauge (11). To express the gravitational field equations in terms of $C_{ij}{}^\a$, it is useful to define the \emph{modified} gravitational field strength
\begin{eqnarray}
\frak{C}_{ij}{}^\a :=\frac 12\,
    C_{ij}{}^\a -C^\a{}_{[ij]}+2e_{[i}{}^\a C_{j]\g}{}^\g\,.\label{12}
\end{eqnarray}
Furthermore, we have
\begin{equation}\label{13} 
{\cal H}^{ij}{}_\a\stackrel{*}{=}\frac{\sqrt{-g}}{8 \pi G}\frak{C}^{ij}{}_\a\,.     
\end{equation}
The GR$_{||}$ field equations are then given by
\begin{equation}\label{14}
 \partial_{[i} C_{jk]}{}^\a\stackrel{*}{=}0\,,
 \end{equation}
\begin{equation}\label{15}
  \partial_j{\cal H}^{ij}{}_\a-{\cal E}_\a{}^i\stackrel{*}{=}{\cal
    T}_\a{}^i\,,
\end{equation}
where ${\cal T}_\a{}^i = \sqrt{-g}  T_\a{}^i$, $T_\a{}^i$ is the energy-momentum tensor and ${\cal E}_\a{}^i$ is defined by
\begin{equation}\label{16}
{\cal E}_\a{}^i:=-\frac 14  e^i{}_\a( C_{ jk}{}^\b
{\cal H}^{jk}{}_\b) + C_{\a k}{}^\b {\cal H}^{ik}{}_\b\,.
\end{equation}

The gravitational field equations in this framework have an interesting analogy with electromagnetic field equations due to the fact that both are gauge theories of Abelian groups. If we first take ${\cal E}_\a{}^i$ in equation (15) to the right-hand side and then ignore for the moment the index  ``$\alpha$'' in equations (14) and (15), we see that these equations are just like Maxwell's equations in a medium. That is, $C_{ij}$ is analogous to the electromagnetic field tensor, so that $C_{ij}$ has components $(\mathbf{E}, \mathbf{B})$ and, similarly, ${\cal H}^{ij}$ has components $(\mathbf{D}, \mathbf{H})$, so that ${\cal E}^i$ in effect has the interpretation of current due to the field itself, since the gravitational field equations are nonlinear. Indeed, ${\cal E}_\a{}^i$ is the energy-momentum tensor density of the gravitational field in GR$_{||}$.
In this interpretation, the tetrad field $e_i{}^\a$, when we ignore ``$\alpha$'', is the analog of the electromagnetic vector potential (i.e., gauge field), except that there are in fact four such potentials here, each for $\alpha = \hat{0},\hat{1},\hat{2},\hat{3}$, as there are four spacetime translations. Let us recall here that GR$_{||}$ is the gauge theory of the four-parameter group of translations in spacetime, whereas electromagnetism is the gauge theory of the one-parameter  $U(1)$ group. Just as in electrodynamics, the homogeneous field equation (14) is automatically satisfied as a consequence of equation (6), while equation (13) is the analog of the constitutive relation for the medium. 

It is important to note that the GR$_{||}$ field equations can also be derived from the action principle
\begin{equation}\label{17}
\d \int\left({\cal L}_{\text{g}}+{\cal L}_{\text{m}} \right)d^4x = 0,
\end{equation}
where ${\cal L}_{\rm g}$ and ${\cal L}_{\rm m}$ are  the gravitational and matter Lagrangian densities, respectively. Moreover, in analogy with electrodynamics, ${\cal L}_{\rm g}$ is quadratic in the gravitational field strength, so that
\begin{equation}\label{18}
{\cal L}_{\rm g}=\frac 12 C_{ij}{}^\a\frac{\partial{\cal L}_{\rm g} }
{\partial C_{ij}{}^\a}=-\frac 14{\cal H}^{ij}{}_\a C_{ij}{}^\a\,
\end{equation}
and $\d {\cal L}_{\rm m}/\d e_i{}^\a={\cal T}_\a{}^i$. Further aspects of the GR$_{||}$ field equations can be found, for instance, in~\cite{28} and~\cite{49}.

The proof that in the classical domain this theory is indeed equivalent to GR is given, for instance, in~\cite{28}.

 \section{Nonlocal GR$_{||}$}
 We recall that in the phenomenological electrodynamics of media, it is possible for the constitutive relation to be nonlocal; for instance, one can have
 \begin{equation}\label{19}
  \mathbf{D}(t)= \mathbf{E}(t) + \int_{-\infty}^{t} f(t-t') \mathbf{E}(t') dt' \,,
 \end{equation}
 where $f$ depends on the properties of the medium---see Eq. (58.3) of~\cite{51}. Nonlocal electrodynamics of accelerated systems has a similar formulation~\cite{47}. This correspondence suggests a natural nonlocal extension of GR$_{||}$, namely, that we should maintain the field equations (14) and (15), but replace the local ``constitutive'' relation (13) with its nonlocal generalization in Weitzenb\"ock spacetime. To this end, we must postulate the existence of a ``constitutive'' kernel that would be responsible for the nonlocal nature of the gravitational interaction. The simplest possibility would be a \emph{scalar} kernel. Moreover, we must ensure that the nonlocal extension of equation (13), once it is fully expressed in terms of holonomic indices, would be consistent with general coordinate invariance. This can be done by means of the world function $\Omega$ and its covariant derivatives~\cite{52}. Let us assume that there exists a \emph{unique} geodesic joining every pair of events $x$ and $x'$ in the spacetime region of interest; then, $\Omega$ is a smooth biscalar in $(x, x')$ such that $2 \Omega$ is the square of the invariant geodesic interval between $x$ and $x'$. It satisfies the basic partial differential equations
\begin{equation}\label{20}
2\Omega=g^{ab}\Omega_a\Omega_b=g^{ij}\Omega_i\Omega_j\,,
\end{equation}
so that the vectors $\Omega_a(x,x'):={\partial\Omega}/{\partial x^a}\,$ and
$\Omega_i(x,x'):={\partial\Omega}/{\partial x'^i}\,$ are tangents to the geodesic at $x$ and $x'$, respectively, and have the same length as the geodesic. Here indices $a, b, c, ...$ refer to $x$ and $i, j, k, ...$ refer to $x'$. Differentiating equation (20), we find that $\Omega_{ai}(x,x')=\Omega_{ia}(x,x')$ and these tend to $-g_{ai}(x)$ in the limit where $x$ and $x'$ coincide. Let us recall here that in Minkowski spacetime, 
\begin{equation}\label{21}
\Omega(x,x')=\frac 12\eta_{ij}(x'^i-x^i)(x'^j-x^j)\,,
\end{equation}
so that $\Omega^a = x^a - x'^a$, $\Omega^i = x'^i - x^i$, $\Omega_{ai}=-\eta_{ai}\,$, $\Omega_{ab}=\eta_{ab}\,$ and $\Omega_{ij}=\eta_{ij}\,$. Further properties of the world function can be found in~\cite{52} and~\cite{49}. A possible nonlocal generalization of equation (13) is then
\begin{eqnarray}\label{22}
  {\cal H}^{ab}{}_c(x)&\!\!\stackrel{*}{=}\!\!&\frac{\sqrt{-g(x)}}{8 \pi G}\,
  \lbrack\frak{C}^{ab}{}_c(x)
    -\int U(x,x')\Omega^{ai}\Omega^{bj}\Omega_{ck} \,
  \nonumber\\ &&
\times \hat{K}(x,x')
    \frak{C}_{ij}{}^k(x')\sqrt{-g(x')}\,d^4 x'\rbrack\,.
\end{eqnarray}
Here the function $U$ has been introduced to ensure that equation (22) is consistent with causality; that is, $x'$ must be in the past of $x$. Thus $U = 0$ if $x'$ is in the future of $x$; otherwise, $U = 1$. 

It remains to discuss the nature of our nonlocal scalar kernel $\hat{K}(x,x')$. It could depend upon $\Omega$ as well as spacetime scalars formed from its covariant derivatives such as, for instance, $\Omega^a e_a{}^\a(x)$ and $\Omega^i e_i{}^\a(x')$. Moreover, $\hat{K}$ could depend on the invariants constructed from the gravitational field strength $C_{ijk}$. As a third rank tensor that is antisymmetric in two indices, $C_{ijk}$ has three 
irreducible pieces: a tensor, a vector and an axial vector. Out of these, one 
can build  three linearly independent quadratic invariants that
 turn out to be equivalent to the three Weitzenb\"ock invariants
\begin{equation}\label{23}
C_{ijk}C^{ijk}\,,\quad C_{kji}C^{ijk}\,,\quad C_{ij}{}^j C^{ik}{}_k
\end{equation}
at $x$ and $x'$. Of course, one could also imagine various combinations of such possibilities.  

The substitution of equation (22) in equation (15) results in the main nonlocal gravitational field equation. The analysis of the physical consequences of the resulting nonlinear and nonlocal theory  is a daunting task; therefore, we resort to the linear approximation in the rest of this paper.

\section{Linear Approximation}

The linear approximation in GR$_{||}$ is defined by
\begin{equation}\label{24}
 e^i{}_\a=\d^i _\a -\psi^i{}_\a\,,\quad  e_i{}^\a={\d}_i ^\a+\psi^\a{}_i\,,
\end{equation}
where $\psi_{i\a}$ is assumed to be small and will be treated to first order, in which case the distinction between holonomic and anholonomic indices disappears. It follows that
\begin{equation}\label{25}
g_{ij}=\eta_{ij}+h_{ij}\,,\qquad h_{ij}=2\psi_{(ij)}\,,
\end{equation}
so that the gravitational field strength is given by
\begin{equation}\label{26} 
C_{ij}{}^k=2\psi^k{}_{[j,i]}\,,
\end{equation}
\begin{eqnarray}
\nonumber\frak{C}^{ij}{}_k&=&-\frac
12\left(h_{k\,\;,}^{\,\;i\;\,j}-h_{k\,\;,}^{\,\;j\;\,i}\right)+\psi^{[ij]}{}_{,k}\\
\label{27}&&+\, \d^i_k\left(\psi_,{}^j-\psi_{l\;\,,}^{\;\,j\;
    \,l}\right)-\d^j_k\left(\psi_,{}^{\,i}- \psi_{l\;\,,}^{\,\;i\;\,l}\right)\,,
\end{eqnarray}
where $\psi=\eta_{ij}\psi^{ij}$.

Let us introduce the trace-reversed potentials $\overline{h}_{ik}$ such that
 \begin{equation}\label{28}
\overline{h}_{ik}=h_{ik}-\frac 12\eta_{ik}h\,,\quad h=\eta_{ij}h^{ij}=2\psi\,.
\end{equation} 
Then,
\begin{equation}\label{29}
\partial_j\frak{C}_{i\;{}k}^{\;j}=-\frac
12\square\, 
\overline{h}_{ki}+\frac 12\overline{h}_{k\,\;,ij}^{\,\;j}+\frac
12\overline{h}_{i\;,kj}^{\,\;j}-\frac
12\eta_{ki}\overline{h}^{jl}{}_{\,\,,jl}\,,
\end{equation}
which is simply the symmetric Einstein tensor $G_{ik}$ in the linear approximation. We now need to substitute the linearized form of the nonlocal ``constitutive'' relation (22) in the linearized version of equation (15). Therefore, the nonlocal field equation can be written in this approximation as
\begin{equation}\label{30}
G_{ik}(x) + \eta^{ab}\int\frac{\partial
{K}(x,y)}{\partial
x^a}\frak{C}_{ibk}(y)d^4y= 8 \pi G T_{ki}(x)\,,
\end{equation}
where $K(x, x') := U(x, x')\hat{K}( x, x')$ and $\partial_iT^{ji} = 0$. Let us note that in the absence of the nonlocal kernel, we recover the linear approximation to general relativity. 
In general, $T_{ij}$ is not symmetric; therefore,  in equation (30) we have 16 field equations for the 16 components of $\psi_{ij}$. 

\section{Newtonian Limit}

The Newtonian limit in the present framework corresponds to the assumption that 
\begin{equation}\label{31}
\psi_{00}=\psi_{11}=\psi_{22}=\psi_{33}=\frac{1}{c^2}\Phi\,, \quad \psi=-\frac{2\Phi}{c^2}\,,
\end{equation}
where $\Phi$ is the Newtonian gravitational potential and the other (off-diagonal) components of $\psi_{ij}$ are neglected in this limit. This is related to the fact that in the Newtonian regime of GR, $g_{ij}=\eta_{ij}+h_{ij}$, where
$h_{ij}=2c^{-2}\Phi\,$ diag$(1,1,1,1)$. We therefore have $c^2 \frak{C}_{0A0}=-2\partial_A\Phi$, where $A, B, C, ...$ here indicate spatial indices that range over 1, 2 and 3. Moreover,  $c^2G_{00}=2\nabla^2\Phi$ as in GR and $T_{00}=\rho c^2$, so that equation (30) reduces to 
\begin{equation}\label{32}
  \nabla^2\Phi(\mathbf{x})+\sum_A\int\frac{\partial
    k(\mathbf{x},\mathbf{y})}{\partial x^A}\frac{\partial\Phi
    (\mathbf{y})}{\partial y^A}d^3y=4\pi G\rho(\mathbf{x})\,,
\end{equation}
where 
\begin{equation}\label{33}
K(x,y)=\delta(x^0 - y^0) k(\mathbf{x},\mathbf{y})\,,
\end{equation}
since all retardation effects are neglected in the Newtonian limit ($c\to\infty$).

The kernel $k(\mathbf{x},\mathbf{y})$ in this nonlocal modification of Poisson's equation of Newtonian gravity can depend on $\mathbf{x}-\mathbf{y}$ as well as on the Weitzenb\"ock invariants at $\mathbf{x}$ and $\mathbf{y}$. The latter have been calculated in this case~\cite{50} and the result is essentially $(\nabla\Phi)^2$. Thus the kernel can be written as
\begin{equation}\label{34}
  k(\mathbf{x},\mathbf{y})=k'(\mathbf{x}-\mathbf{y})+k''\left(\mathbf{x}
    -\mathbf{y};\frac{|\nabla_{\mathbf{y}}\Phi|}{|\nabla_{\mathbf{x}}
      \Phi|}\right)\,.
\end{equation}
The second term in equation (34) involves a nonlinear modification of Newtonian gravity. To simplify matters, we concentrate on the first term in what follows. The resulting \emph{linear} nonlocal modification of Poisson's equation is then
\begin{equation}\label{35}
  \nabla_{\mathbf{x}}^2\Phi(\mathbf{x})+\int k'(\mathbf{x}-\mathbf{y})
\nabla_{\mathbf{y}}^2\Phi(\mathbf{y}) d^3y  =4\pi G \rho(\mathbf{x})\,.
\end{equation}
Using the standard Liouville-Neumann method of successive substitutions, equation (35) can be expressed as
\begin{equation}\label{36}
\nabla^2\Phi=4\pi G ( \rho + \rho_D)\,, 
\end{equation}
where
\begin{equation}\label{37}
\rho_D(\mathbf{x})=\int q(\mathbf{x}-\mathbf{y})\rho(\mathbf{y})d^3y \,.
\end{equation}
Here $q$ is a universal function that is independent of the nature of the source and is the reciprocal  kernel to $k'$~\cite{48,49,50}. Thus nonlocality can appear as an extra matter source (``dark matter'') in the standard local theory. 

It is interesting to see if the nonlocal theory can actually provide a natural solution to the dark matter problem in astrophysics~\cite{53,54,55}. To this end, we consider a spiral galaxy of mass $M$ and, for simplicity, let $\rho(\mathbf{x}) = M \delta(\mathbf{x})$ and
\begin{equation}\label{38}
\rho_D(\mathbf{x})=\frac{v_0^2}{4 \pi G}\frac{1}{|\mathbf{x}|^2}
\end{equation}
be the density of dark matter, with corresponding asymptotic speed $v_0$, that can account for the flat rotation curves of spiral galaxies. Then it follows from equation (37) that 
\begin{equation}\label{39}
q(\mathbf{x}-\mathbf{y})=\frac{1}{4\pi\lambda}
\frac{1}{|\mathbf{x}-\mathbf{y}|^2}\,,
\end{equation}
where $\lambda=GM/v_0^2$ is a constant of order 1\,kpc for spiral galaxies. The solution of  the modified Poisson equation in this case is
\begin{equation}\label{40}
\Phi(\mathbf{x})=-\frac{GM}{|\mathbf{x}|}
+\frac{GM}{\lambda}\ln\left(\frac{|\mathbf{x}|}{\lambda}\right)\,.
\end{equation}
It turns out that such a modification of Newtonian gravity was first phenomenologically proposed by Tohline~\cite{56} and later developed by Kuhn and his coworkers~\cite{57,58} in order to account for the dark matter problem in spiral galaxies and galaxy clusters. In the Tohline-Kuhn scheme, $\lambda$ is a universal constant and hence $M \propto v_0^2$; however, the empirical Tully-Fisher relation~\cite{59,60} implies that $M \propto v_0^4$ is observationally preferred. The inclusion of nonlinearity in the modified Poisson equation via an appropriate $k''$ in equation (34) may solve this problem. 

The implications of modified Newtonian potential (40) for gravitational physics in the solar system are briefly studied in~\cite{50}. It is important to look for any observational evidence of nonlocal gravity in the solar system; however, at present such effects appear to be too small to be detectable~\cite{50}.

\section{Discussion}
Starting from first principles, we have presented a framework for nonlocal gravity that has the potential to provide a \emph{natural} theoretical explanation for the observational evidence regarding dark matter. Indeed, nonlocality appears to simulate dark matter by introducing in effect a new source term (``dark matter'') in Einstein's theory of gravitation. In the Newtonian regime, the density of dark matter is essentially the convolution of matter density with a kernel that represents the novel nonlocal aspect of the gravitational interaction and must ultimately be determined from observation. A more detailed treatment of nonlocal gravity is contained in~\cite{48,49,50}.


\begin{theacknowledgments}
  This paper is based on lectures delivered at the XIV Brazilian School of Cosmology and Gravitation (August 30 - September 11, 2010). I thank M\'ario Novello and the organizing committee for their kind invitation and excellent hospitality.  I am grateful to Friedrich Hehl for helpful discussions on all aspects of nonlocal gravity.
\end{theacknowledgments}

\bibliographystyle{aipproc}



\bibliographystyle{aipproc}   

\bibliography{sample}

\IfFileExists{\jobname.bbl}{}
 {\typeout{}
  \typeout{******************************************}
  \typeout{** Please run "bibtex \jobname" to optain}
  \typeout{** the bibliography and then re-run LaTeX}
  \typeout{** twice to fix the references!}
  \typeout{******************************************}
  \typeout{}
 }

\end{document}